\documentclass[9pt,twocolumn,twoside]{osajnl}

\journal{ol} % Choose journal (ao, aop, josaa, josab, ol)

\setboolean{shortarticle}{true} 
\def\omegab{\bar{\omega}}

\ifthenelse{\boolean{shortarticle}}{\colorlet{color2}{color2b}}{\colorlet{color2}{color2}} % Automatically switches colors for short articles

\title{Photon Pair Generation with Tailored Frequency Correlations in Graded-Index Multimode Optical Fibers}

\author[1,2]{Hamed Pourbeyram}
\author[1,2,*]{Arash Mafi}

\affil[1]{Department of Physics \& Astronomy, Univ. of New Mexico, Albuquerque, NM 87131, USA}
\affil[2]{Center for High Technology Materials, Univ. of New Mexico, Albuquerque, NM 87106, USA}

\affil[*]{Corresponding author: mafi@unm.edu}

\dates{Compiled \today}

\ociscodes{(270.0270) Quantum optics; (270.5585) Quantum information and processing; (190.4380) Nonlinear optics, four-wave mixing; (060.4370) Nonlinear optics, fibers}

\doi{}
\begin{abstract}
We study theoretically the generation of photon pairs with controlled spectral correlations via the four-wave mixing (FWM) process
in graded-index multimode optical fibers (GIMFs). We show that the quantum correlations of the generated photons in GIMFs can be 
preserved over a wide spectral range for a tunable pump source. Therefore, GIMFs can be utilized as quantum-state-preserving tunable 
sources of photons. In particular, we have shown that it is possible to generate factorable two-photon states, which allow for 
heralding of single-photon pure-state photons without the need for spectral post filtering. We also elaborate on the possibility of 
simultaneously generating correlated and uncorrelated photon pairs in the same optical fiber.
\end{abstract}

\setboolean{displaycopyright}{true}

\begin{document}
\maketitle
\thispagestyle{fancy}
\ifthenelse{\boolean{shortarticle}}{\ifthenelse{\boolean{singlecolumn}}{\abscontentformatted}{\abscontent}}{}
%%%%%%%%%%%%%%%%%%%%%%%%%%%%%%%%%%%%%%%%%%%%%%
The ability to control the coherent dynamics of photons and correlations among them paves the way to generate 
a wide range of quantum states from heralded single photons to entangled photon pairs. FWM 
in optical fibers has been used to spontaneously generate correlated photon pairs~\cite{2005-Kumar-Li-1,2005-Russell-Rarity,2006-Agrawal-Lin-1,
2008-Migdall-Goldschmidt-1,2009-Walmsley-Cohen,2014-Lorenz-Fang-2,2014-URen-Cruz-Delgado,2016-Raymer-Smith,2016-Mosley-Francis-Jones}. 
These signal-idler photons, sometimes referred to as {\em daughter} photons, share some information due to the 
conservation of energy and momentum. 
The photon pair correlations can be manipulated via tailoring the joint spectral amplitude (JSA) of the photon 
pair~\cite{2001-Walmsley-Grice,2007-Walmsley-GarayPalmett,2009-Walmsley-Cohen,2016-Rottwitt-Christensen,2012-Zhao,2016-Mafi-Pourbeyram-PRA}, giving access to a 
variety of quantum states, from a highly correlated photon pair (entangled)~\cite{2004-Kumar-Li} to an uncorrelated pair (factorable)
that can be used as a heralded pure-state single photon source~\cite{2011-Walmsley-Soller,2013-Lorenz-Fang}.

The extra degrees of freedom available in multimode optical fibers can be used to control the physical attributes of
the photon pairs generated via the spontaneous FWM process. The presence of multiple spatial modes allows for the 
intermodal FWM (IM-FWM) process, which can result in signal and idler photons that have large spectral separations 
from the pump~\cite{1981-Lin,2015-Mafi-Pourbeyram-OPEX,2017-Dupiol-Kurpa}; therefore, they are minimally contaminated by the scattered and 
residual pump and Raman photons. Different spatial mode combinations can also
result in different IM-FWM processes that allow for the simultaneous generation of multiple photon pairs in the 
same fiber~\cite{2014-URen-Cruz-Delgado,2016-Mafi-Pourbeyram-PRA}. The diversity of the group velocities of different spatial modes can also 
be used as an effective tool to manipulate the JSA of the photons pairs.  

Among the general class of MMFs, GIMFs exhibit unique dispersive, nonlinear, and 
spatiotemporal properties~\cite{2012-Mafi,2015-Wise-Wright,2016-Agrawal-Buch,2017-Wise-Wright}. For example, their modes can be
classified into mode-groups with identical intra-group and 
equally-spaced inter-group phase velocities~\cite{2012-Mafi,2016-Mafi-Nazemosadat}. All modes can also be designed to have nearly identical group velocities 
near a special wavelength~\cite{2012-Mafi}. Therefore, short high peak power laser pulses do not easily disintegrate due to intermodal 
group velocity dispersion and the laser pulses go through a rapid submillimeter-length self-imaging pattern when propagating 
along these fibers~\cite{2011-Mafi-OL}. These unique properties make GIMFs an appealing platform for observing novel nonlinear optic phenomena.  
Examples include the observation of multimode solitons~\cite{2015-Wise-Wright-OPEX,2016-Wise-Zhu}, supercontinuum generation~\cite{2016-Christodoulides-LopezGalmiche,2016-Kurpa-Wabnitz},
multimode saturable absorption~\cite{2013-Mafi-Nazemosadat}, self-induced beam cleanup~\cite{2016-Christodoulides-LopezGalmiche,2013-Mafi-Hamed}, 
and geometric parametric instability~\cite{2003-Longhi,2016-Krupa-PRL,2017-Teugin}. In this paper, we leverage the unique 
dispersive and nonlinear properties of GIMFs for the generation of photon-pair states with a high degree of control over their 
spectral correlations for an ultra-broad spectral range. 

We recently explored the tailoring of the JSA of the photon pair generated in a commercial 
multimode step-index optical fiber via the IM-FWM process~\cite{2016-Mafi-Pourbeyram-PRA}. We showed that it is possible to generate factorable 
two-photon states exhibiting minimal spectral correlations between the photon pair components in conventional multimode fibers 
using commonly available pump lasers. In this paper, we extend our studies to the case of photon pair generation in GIMFs.
We show that a commercial GIMF can be used as a robust medium to generate factorable photon pairs over a large bandwidth
that can be considerably larger than that obtained in step-index fibers. This interesting property is rooted in the special dispersive
attributes of GIMFs.

We first present a brief overview of GIMFs and establish the notation that will be used in the rest of the paper.
The refractive index profile of a GIMF is given by
%%%%%%%%%%%
\begin{equation}
\label{eq:indexprofile}
n^2(\rho)=n_0^2\left[1-2\Delta\left(\frac{\rho}{R}\right)^\alpha\right].
\end{equation}
%%%%%%%%%%%
$R$ is the core radius, $n_0$ is the refractive index in the center of the core,
$\Delta$ is the relative index difference between the core and cladding, $\alpha\approx 2$ characterizes a
near parabolic-index profile in the core ($\rho\le R$), and $\alpha = 0$ in the cladding ($\rho>R$).

The guided linearly polarized (LP) modes are represented by Laguerre-Gaussian 
functions~\cite{2012-Mafi}. Each mode is identified by two integers, $p$, and $m$, referred to as the radial and angular numbers, respectively. 
In this notation, each mode is shown in the form $LP_{|m|,p+1}$. The mode-group number is also defined as $g=2p+|m|+1$.
All the guided modes with an identical mode-group number $g$ are nearly degenerate in the value of their propagation constant $\beta_g$~\cite{2012-Mafi,2016-Mafi-Nazemosadat} defined by
%%%%%%%%%%%%%%%%
\begin{align}
\label{Eq:beta}
&\beta_g=n_0k \left(1-2N \Delta \right)^{1/2},\\
&N=\left(\frac{g}{\sqrt{N_\alpha}}\right)^{2\alpha/(\alpha+2)},\quad N_\alpha=\frac{\alpha}{\alpha+2}n_0^2k^2R^2\Delta\,.
\end{align}
%%%%%%%%%%%%%%%%
$N_\alpha$ indicates the total number of guided modes (considering the polarization degeneracy), and $k=2\pi/\lambda$, where $\lambda$ is the wavelength. 
For the calculations in this paper, we have considered a commercial GIMF with typical parameters which can be found in 
Refs.~\cite{2012-Mafi,2016-Mafi-Nazemosadat}. Briefly, $R=25\,\mu m$, $\Delta=0.01$, and $\alpha\approx 2.11$.

The phase-matching for IM-FWM is given by
%%%%%%%%%%%%%%%%%%
\begin{equation}
\label{eq:phasematching1}
\beta_\nu^{p}+\beta_\kappa^{p}-\beta_\mu^{s}-\beta_\xi^{i}=0, 
\end{equation}
%%%%%%%%%%%%%%%%%%
where the superscripts signify the pump (p), signal (s), and idler (i). The subscripts index the specific Laguerre-Gaussian spatial mode
in which the pump, signal, or idler propagate. Because the propagation constants only depend on the mode-group numbers (see Eq.~(\ref{Eq:beta})),
the phase matching can be readily cast into a relationship among the mode-group numbers \{$g^{(1)}_p$, $g^{(2)}_p$, $g_i$, $g_s$\}, representing 
the two pump photons, the idler photon, and the signal photon, respectively.

The JSA of the signal-idler photon pair generated in a optical fiber can be constructed~\cite{2007-Walmsley-GarayPalmett,2009-Walmsley-Smith-1,2016-Mafi-Pourbeyram-PRA} using 
the group delay between pump-signal and pump-idler photons given by
%%%%%%
\begin{align}
\tau_j&=L\left[\beta^{(1)}_p(\omegab_p)-\beta^{(1)}_j(\omegab_j)\right],\quad  j\in\{s,i\},
\label{Eq:tau}
\end{align}
%%%%%%
and the pump bandwidth $\sigma_p$. $L$ is the fiber length. In Eq.~(\ref{Eq:tau}) $\beta^{(1)}$ is the inverse of the group 
velocity defined as $\partial\beta/\partial\omega$, and $\omegab_j ~(j\in\{p,s,i\})$ indicate the 
phase-matched frequencies. Using $\tau_s$, $\tau_i$, and $\sigma_p$, it is possible to determine the 
correlation among signal and idler by calculating the purity of the quantum state, $\mathcal{P}$, for idler (or signal) photon. 
The explicit form of $\mathcal{P}$ is given by~\cite{2016-Mafi-Pourbeyram-PRA}
%%%%%%
\begin{equation}
\mathcal{P}=\sqrt{\frac{2\,\eta\,r_2^2\,(X_1-2)X_1}{\Big(2\,r_2^2\,X_1+\eta\, r_1\Big)\Big(2\,r_2^2\,X_1+\eta/r_1\Big)}},\qquad \eta\approx 0.193,
\label{eq:finalPurity}
\end{equation}
%%%%%%
which is parameterized using two dimensionless parameters $r_1$ and $r_2$ defined as
%%%%%%
\begin{equation}
r_1=\frac{\tau_s}{\tau_i}, \quad r_2=\frac{\sigma_{si}}{\sigma_p},\quad \sigma_{si}=\dfrac{1}{\sqrt{\tau_s^2+\tau_i^2}},
\quad X_1=r_1+\dfrac{1}{r_1}.
\end{equation}
%%%%%%
Note that $\mathcal{P}$ ranges from unity to zero, where a pure quantum state ($\mathcal{P}=1$) indicates a spectrally uncorrelated signal-idler photon pair, 
giving  a factorable quantum state.

%%%%%%
\begin{figure}[htpb]
\includegraphics[width=3.2 in]{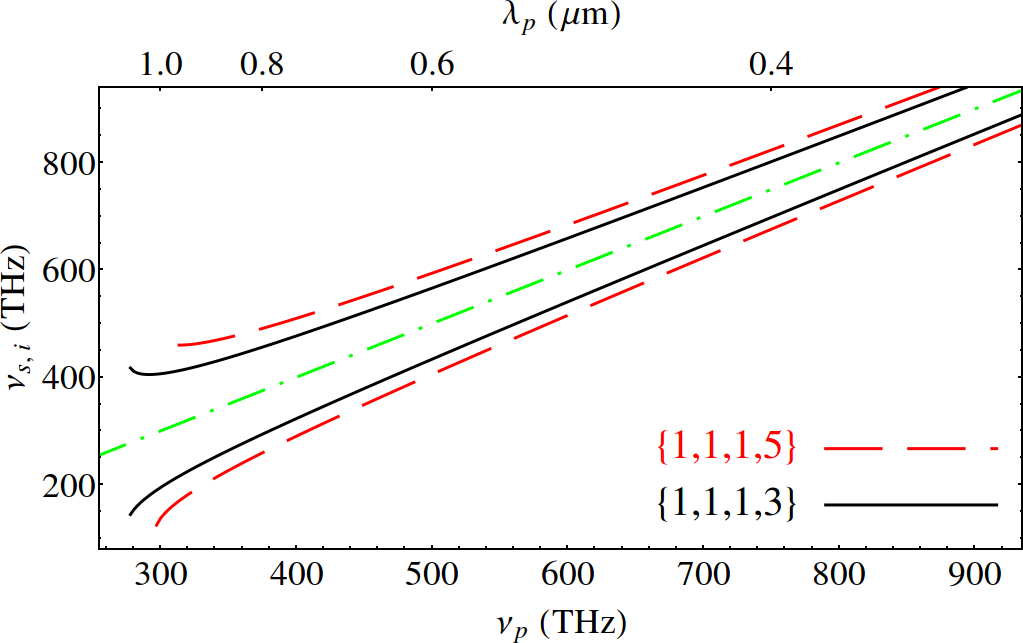}
\caption{Phase-matched FWM frequencies $\nu_s,\nu_i$ as a function of pump frequency $\nu_p$ ($\nu_j=\omega_j/2\pi$), for two distinct FWM processes.
The dot-dashed green line represents the pump line.}
\label{Fig:pm}
\end{figure}
%%%%%%
In Fig.~\ref{Fig:pm} we show the phase-matched signal-idler frequencies ($\nu_s$ and $\nu_i$) as a function of pump frequency ($\nu_p$), 
for two FWM processes in the GIMF. In this figure, $\{g^{(1)}_p,g^{(2)}_p,g_i,g_s\}=\{1,1,1,3\}$ is labeled for example as $\{1,1,1,3\}$.
There is clearly a linear dependence of $\omegab_s$ ($\omegab_i$) versus $\omegab_p$ over a wide range of pump spectrum for all depicted processes.
It can be generally shown that~\cite{2016-Mafi-Pourbeyram-PRA,Offir-Cohen}
%%%%%%
\begin{equation}
r_1=1-2\,\dfrac{d\omegab_p}{d\omegab_s};
\label{Eq:r1}
\end{equation}
%%%%%%
therefore, according to Eq.~(\ref{Eq:r1}), if $d\omegab_p/d\omegab_s$ does not change with varying the pump wavelength (as is the case for a broad spectral range in Fig.~\ref{Fig:pm}), 
the value of $r_1$ does not change with the pump wavelength. Therefore, the value of purity can be easily kept constant by keeping $r_2$ unchanged 
by merely rescaling the length of the fiber $L$. Later, we will show that $r_1\approx -1$ is easily accessible for GIMFs; therefore,
one can generate an ultra-broadband tunable photon pair source with $\mathcal{P}\approx 1$ by the judicious selection of $L$ or $\sigma_p$. 

%%%%%%
\begin{figure}[htpb]
\includegraphics[width=3.5 in]{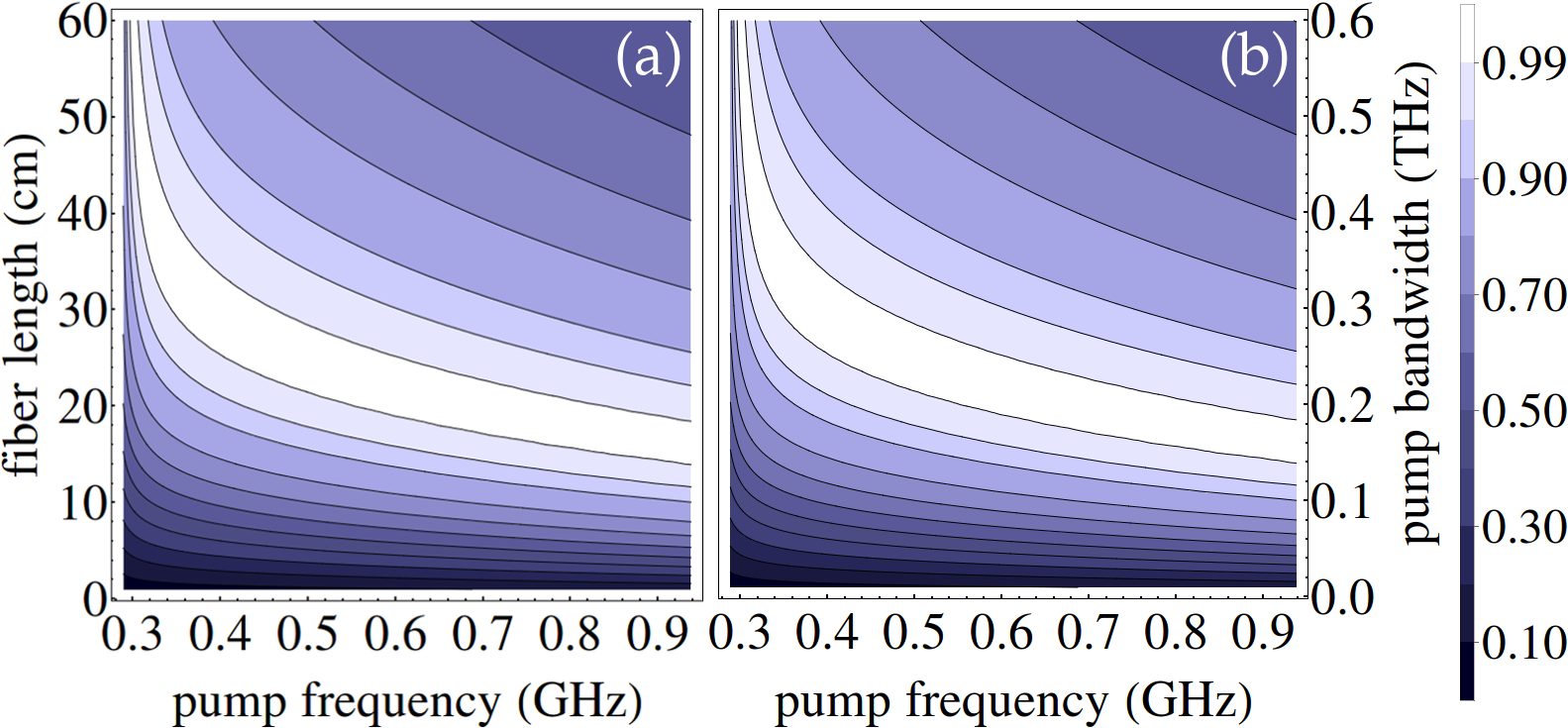}
\caption{Purity contours as a function of pump frequency $\nu_p$ vs. (a) fiber length for fixed pump bandwidth of 0.5 THz, and (b) pump bandwidth for a 50 cm long fiber. Both contours are for
\{1,1,1,3\} process.}
\label{Fig:contour}
\end{figure}
%%%%%%
In Fig.~\ref{Fig:contour}, we demonstrate our theoretical calculation of purity as a function of the pump frequency versus the (a) fiber length, and (b) pump bandwidth. 
The calculations are performed for the \{1,1,1,3\} process.
In Fig.~\ref{Fig:contour}(a), it can be seen that for the fixed value of $\sigma_p$, $\mathcal{P}\approx 1$ can be achieved by tuning the value of $L$.
Moreover, even if the value of $L$ is kept fixed, it is possible to maintain high purity over a reasonably large pump frequency range, making it highly desirable
from a device standpoint. A similar argument can be made in Fig.~\ref{Fig:contour}(b) by reversing the role of $L$ and $\sigma_p$.   

The results presented in Fig.~\ref{Fig:contour} are for the \{1,1,1,3\} process. In the following, we will show that the value of purity in GIMFs,
in general, depends mainly on the spectral separation of the signal/idler from the pump ($\Omega$). This is an important result from an experimental standpoint
because one will not even need to examine the mode profiles and can evaluate the value of the purity expected from the process only by a simple
spectral measurement. From a theoretical standpoint, this result is interesting because the frequency separation is analytically calculable 
knowing the mode-group numbers involved in the FWM process~\cite{2016-Mafi-Nazemosadat}. 

In order to show that the purity in GIMFs depends mainly on $\Omega$, we expand Eq.~(\ref{Eq:beta})
to the first order in the small parameter $\Delta$ and obtain
%%%%%%%%%%%%
\begin{equation}
\label{eq:betaapprox}
\beta_g\approx n_0(\lambda)k -\dfrac{\sqrt{2\Delta}}{R}g.
\end{equation}
%%%%%%%%%%%%
Because terms of higher orders in $\Delta$ have been ignored, this approximation is only reliable if the length of the fiber $L$ satisfies
$L\ll L_0/g^2$, where $L_0=\pi n_0kR^2/\Delta$~\cite{2012-Mafi}. For the commercial GIMF in Refs.~\cite{2012-Mafi,2016-Mafi-Nazemosadat}, we have $L_0\approx 2.2$\,m
at $\lambda=850$\,nm. Therefore, this approximation holds for the typical fiber lengths considered in this paper for photon pair generation. 
Using Eq.~(\ref{eq:betaapprox}) and ignoring the frequency dependence of $\Delta$, we have $\beta^{(1)}=\partial_{\omega}(n\omega/c)$; therefore, the 
group velocity is independent of the group number $g$. Furthermore, we can expand the  $\beta^{(1)}$ belonging to the signal (idler) around the 
$\omegab_p$ and only keep the terms up to the first order in the signal/idler-pump frequency separation $\Omega$ to get even a more simplified equation for the group delay 
$\tau$. We will discuss the validity of these approximations later in this paper. In this case, the group delays can be written as
%%%%%%
\begin{equation}
\tau_s\approx+\Omega~\frac{L}{c}~\partial^{2}_{\omega}(n\omega)|_{\omegab_p}~,\quad \tau_i\approx-\Omega~\frac{L}{c}~\partial^{2}_{\omega}(n\omega)|_{\omegab_p}~.
\label{Eq:tau-approx}
\end{equation}
%%%%%%

%%%%%%
\begin{figure}[t]
\includegraphics[width=3.4 in]{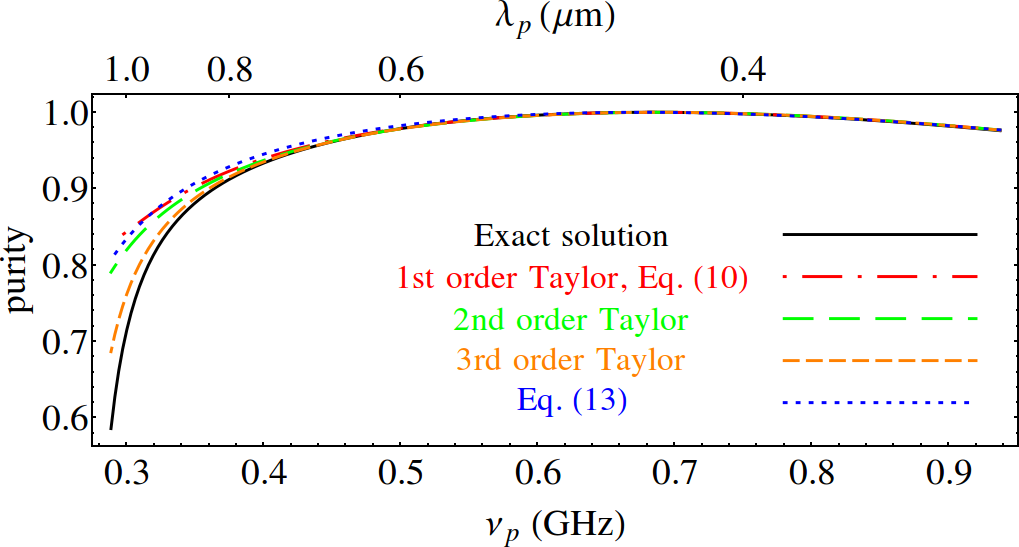}
\caption{Purity vs. pump frequency using different methods discussed. The spectral separation is for \{1,1,1,3\} process, G=2, fiber length is 50 cm, and pump bandwidth is 0.2 THz.}
\label{Fig:P-comparison}
\end{figure}
%%%%%%
In GIMFs, $\Omega$ can be accurately estimated using the material dispersion and physical parameters of the fiber by~\cite{2016-Mafi-Nazemosadat}
%%%%%%%%%%%%
\begin{equation}
\label{eq:freqshift}
\Omega^2 \approx \dfrac{\sqrt{2\Delta}}{R}\dfrac{G~c}{\partial^{3}_{\omega}(n\omega)},
\end{equation}
%%%%%%%%%%%%
where $G=g_s+g_i-g^{(1)}_p-g^{(2)}_p$. $G$ is determined by the modes involved in the FWM process, e.g. $G=2$ is used for the FWM process identified by \{1,1,1,3\}. 
Using Eqs.~(\ref{Eq:tau-approx}) and ~(\ref{eq:freqshift}),  $r_1$ and $r_2$ simplify to
%%%%%%%%%%%%
\begin{align}
\label{eq:r-approx}
r_1\approx-1~,\qquad
r_2\approx\frac{1}{\sigma_p L}\sqrt{\frac{R~c}{G\sqrt{8\Delta}}}\left[\partial^{2}_{\omega}(n\omega)\right]^{-\frac{1}{2}}~,
\end{align}
%%%%%%%%%%%%
from which the purity is calculated as
%%%%%%%%%%%%
\begin{equation}
\label{eq:purity-approx}
\mathcal{P}=\frac{4~r_2\sqrt{\eta}}{4~r_{2}^{2}+\eta}.
\end{equation}
%%%%%%%%%%%%
We emphasize that in order to use these approximations, it is important for the modes involved in the FWM process to be far from their cutoff frequencies
such that the material dispersion dominates over the waveguide contribution.

In Fig.~\ref{Fig:P-comparison}, we examine the accuracy of the approximations made so far in detail. We consider the purity calculation for the
\{1,1,1,3\} process ($G=2$) for the fiber length of 50\,cm and the pump bandwidth of 0.2\,THz as a function of the pump frequency $\nu_p$.
The solid (black) curve shows the full calculation using Eq.~(\ref{Eq:beta}) and its derivatives, so no approximation is made. The dashed-dotted
(red) curve uses the approximation in Eqs.~(\ref{Eq:tau-approx}), for which the value of $\Omega$ is calculated directly by using Eq.~(\ref{Eq:beta}).
The long-dashed (green) and short-dashed (cyan) curves are calculated in a similar way to the dashed-dotted (red) curve except 2nd and 3rd order Taylor
expansion terms in Eq.~(\ref{Eq:tau-approx}) are retained. The dotted (purple) curve is plotted using the full approximation formula in Eq.~(\ref{eq:purity-approx})
where $\Omega$ is calculated using the approximation in Eq.~(\ref{eq:freqshift}). Therefore, there is a good agreement between the exact solution and these 
approximate methods for a wide range of the pump frequency. Results presented in Fig.~\ref{Fig:P-comparison} also show more clearly the possibility of a 
tunable uncorrelated photon pair source in a GIMF.

%%%%%%
\begin{figure}[t]
\includegraphics[width=3.2 in]{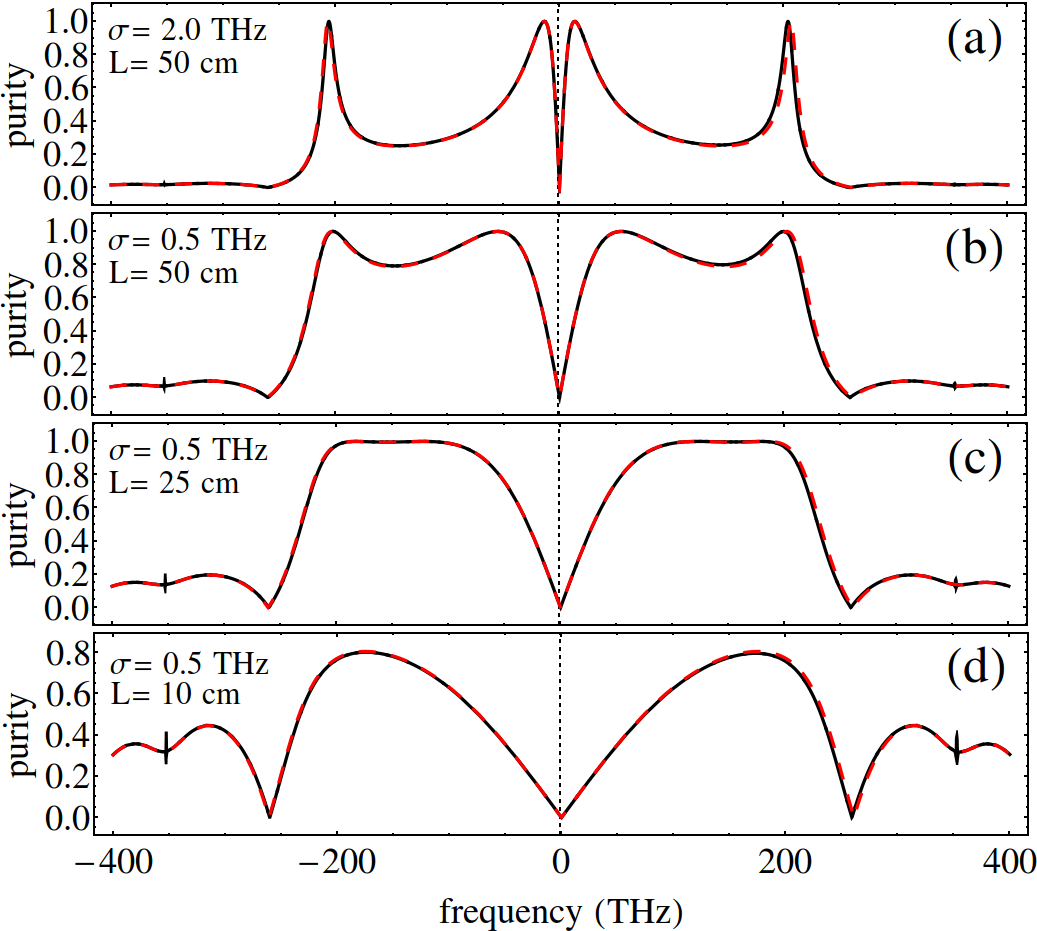}
\caption{Purity as a function of spectral separation from the pump. Dashed red line indicates the result calculated using Eq.~\ref{Eq:tau-approx}, and
the solid black line shows the result calculated from an exact form of the propagation constant in Eq.~(\ref{Eq:beta}) for \{1,1,1,3\}.}
\label{Fig:comparison}
\end{figure}
%%%%%%
According to the approximations presented earlier, we argued that the purity is determined chiefly by the spectral separation $\Omega$, and the dependence 
on the mode-groups numbers appears only indirectly through $\Omega$ and Eq.~(\ref{eq:freqshift}). Therefore, from a practical standpoint in an experiment,
where one usually measures the FWM spectrum rather than the modal content, it is interesting to investigate the purity only as a function of $\Omega$
for various experimental scenarios. This is done in Fig.~\ref{Fig:comparison} where purity is plotted as a function of $\Omega$ for several different values
of $L$ and $\sigma_p$ for the photon pairs generated through the \{1,1,1,3\} process using an 850\,nm wavelength pump. 
In each subfigure, the dashed red line demonstrates the approximate calculation from Eq.~(\ref{Eq:tau-approx}), and the solid black line shows the 
calculation using the exact form of the group velocity. It must be noted that these should be treated as general reference plots; the exact modal content
will fix the value of $\Omega$ at a particular phase-matched frequency shift. The fact that the approximate method (applicable to all values of $G$) and 
the exact method (only applicable to \{1,1,1,3\} though ignoring the phase-matching) give nearly identical results prove the broad applicability of these curves.
Moreover, different scenarios show that by a proper choice of $L$ and $\sigma_p$ it is possible to have high purity over a broad spectral band;
or obtain high purity in one frequency region and low purity in another. Therefore, it is possible to simultaneously generate mixed purity values using
different multimode phase-matching processes in a single strand of optical fiber. 

In conclusion, we have investigated GIMFs as an ultra-broadband source of photon pairs with controlled
spectral correlations. We show that GIMFs can be used as a tunable source of uncorrelated (or correlated) photon pairs 
as their spectral correlations are relatively independent of the pump central wavelength. 
Our calculations indicate that while tuning the pump frequency will change the frequencies of the signal and idler, the 
signal (idler) photon state purity remains unchanged. Therefore, GIMFs can be used to make quantum-state-preserving 
tunable sources of photon pairs. We have also shown that the purity is mainly a function of the spectral separation of the
idler-signal pair from the pump ($\Omega$), and its dependence on the spatial mode content is indirect and only through 
its dependence on $\Omega$. Finally, we have shown that by the right selection of the physical parameters of the system
including the fiber length and/or the pump bandwidth, it is possible to simultaneously generate correlated and uncorrelated 
photon pairs in the same optical fiber.
%%%%%%%%%%%%%%%%%%%%%%%%%%%%%%%%%%%%%%%%%%%%%%
\section*{Acknowledgment}
The authors acknowledge support by the National Science Foundation (NSF) Grant No. 1522933.

%%%%%%%%%%%%%%%%%%%%%%%%%%%%%%%%%%%%%%%%%%%%%%
\providecommand{\noopsort}[1]{}\providecommand{\singleletter}[1]{#1}%

\end{document}